\documentclass[traditabstract]{aa}
\usepackage{natbib}
\usepackage{graphicx}
\bibpunct{(}{)}{;}{a}{}{,}

\begin{document}

\title
{The evolutionary status of the UX Orionis star RZ Piscium}

\author
{V.P.\, Grinin\inst{1,2}\and I.S.\,Potravnov\inst{1}\and
F.A.\,Musaev\inst{3,4}} \institute
{Pulkovo Astronomical Observatory, Russian Academy of Sciences, 196140, Pulkovo, St.\,Petersburg, Russia\\
e-mail: grinin@gao.spb.ru; ilya-astro@yandex.ru \and The Sobolev
Astronomical Institute, St. Petersburg University, Petrodvorets,
St. Petersburg, Russia \and
Special Astrophysical Observatory, Russian Academy of Sciences, Nizhnij Arkhyz, Russia\\
e-mail: faig@sao.ru \and Terskol Branch of the Institute for
Astronomy of RAS, Terskol, Russia}

%\offprints{V.\,P.\,Grinin}

\date{Received/accepted}

\titlerunning{ }

\authorrunning{Grinin et al.}

\abstract {The star RZ Psc is one of the most enigmatic members of
the UX Ori star family. It shows all properties that are typical
for these stars (the light variability, high linear polarization
in deep minima, the blueing effect) except for one: it lacks any
signatures of youth. With the lithium line 6708 \AA\, as a rough
estimate for the stellar age, we show that the "lithium" age of RZ
Psc lies between the age of stars in the Pleiades ($\sim$70 Myr)
and the Orion ($\sim$10 Myr) clusters. We also roughly estimated
the age of RZ Psc based on the proper motion of the star using the
Tycho-2 catalog. We found that the star has escaped from its
assumed birthplace near to the Galactic plane about 30-40 Myr ago.
We conclude that RZ Psc is a post-UXOr star, and its sporadic
eclipses are caused by material from the debris disk.

} \maketitle

\section{Introduction}
The star RZ Psc (Sp = K0 IV, Herbig \cite{Herbig} is one of the
most unusual variable stars. According to the properties of its
light curve, it belongs to the family of young stars of the UX Ori
type, the photometric and polarimetric activity of which is caused
by the variable circumstellar (CS) extinction (Grinin et al.
{\cite{Grin91}) According to the current models (Dullemond et al.
\cite{Dull}), the region of the CS disk close to the dust
evaporation zone, where the main part of the near infrared excess
is formed (Natta et al. \cite{Natta}), is responsible for this
variability. However, RZ Psc shows none of the classical signs of
youth: it has neither infrared (JHK) excess (Glass, \& Penston
\cite{GP74})\footnote{RZ Psc is absent in the IRAS and Spitzer
data bases.}, nor emission in the $H_\alpha$ line (Kaminskii,
Kovalchuk \& Pugach \cite{Kam}). The star lies at the Galactic
latitude $b \simeq -35^\circ$, where there are neither young
stars, nor star-formation regions. The brightness minima of RZ Psc
are also quite unusual: with large amplitudes (up to $\Delta V
\approx$ 2.5 mag.), they are very brief (1-2 days; Zaitseva
\cite{Za85}, Kardopolov, Sahanionok \& Shutemova \cite{KSSh80},
Pugach \cite{Pu81}, Wenzel \cite{Wen89}). Similar minima are
sometimes also observed in another UXOrs but, in the case of RZ
Psc, they are typical. Several attempts were undertaken to find
the period between the star eclipses, but all were unsuccessful
(Wenzel \cite{Wen89}).

The above combination of the conflicting observational properties
poses many questions about the evolutionary status of RZ Psc and
the origin of the eclipses. To clarify these questions, we analyze
the Li 6708 \AA\, spectral line, which is commonly used for age
estimations of solar-type stars (see e.g., Soderblom et al.
\cite{Sod93}; Sestito \& Randich \cite{Ses}; da Silva et al.
\cite{daS}). In addition, we use the proper motion of RZ Psc from
the Tycho-2 Catalog by Hog et al. \cite{Hog} to roughly estimate
the time at which the star escaped from its birthplace near to the
Galactic plane. Both estimates allow us to classify RZ Psc as a
post-UXOr.

\section{The "lithium" age of RZ Psc}
Although RZ Psc has already been investigated for a long time,
most papers have been devoted to the study of its variability.
There is only one paper (Kaminskii et al. 2000) in which the
quantitative analysis of the star spectrum has been made. In that
paper, however, there is no information about the Li 6708 \AA\,
line. To study this line, the spectrum of RZ Psc was obtained at
the Terskol Observatory (Russia). The observations were made on
2009 Nov. 9, with the echelle spectrograph SPECPHOT (spectral
resolving power R = 13500) at the 2 m telescope. The spectrum was
reduced with the software package DECH by Galazutdinov \cite{Gal},
which provides all standard tasks of CCD image and spectra
processing. The wavelength calibration was made with an Fe-Ar
comparison lamp spectrum.

A portion of the reduced spectrum around the Li 6708\, \AA \, line
is displayed in Fig. 1. One can see that this line is present in
the spectrum of RZ Psc and has a moderate depth, with an
equivalent width of \emph{EW}(Li) = 0.202 $\pm$ 0.010 \AA. For
comparison, in Fig.~\ref{Lit} we show two versions of a synthetic
spectrum calculated for the same spectral range with the numerical
code from Piskunov \cite{Pis92}) and the VALD data base by
Piskunov et al. \cite{Pis95}, Kupka et al. \cite{Kup}. The spectra
were broadened by rotation with $v\,\sin{i}$ = 23 km/s (Kaminskii
et al. 2000). In both cases we used the Kurucz \cite{Kur} model of
atmosphere with $T_{ef}$ = 5250 $K$, $\log{g}$ = 4.0, and
$v_{turb}$ = 2 km/s. In one of these spectra, the Li abundance is
the same as in the Sun. In the other one, it is 100 times higher.
We see that in the first case, the Li 6708 \AA\, line is
practically invisible. In the second case, the lithium line in the
synthetic spectrum almost coincides with the observed one. Taking
into account the possible uncertainties in the estimation of the
spectral type of the star, we have calculated several additional
synthetic spectra, changing the model parameters in the ranges
$T_{ef}$ = $\pm$ 250 $K$ and $\log{g}$ = $\pm$ 0.5, and these
calculations confirmed the result obtained above.

\begin{figure}
%\begin{centering}
%\label{1}
\includegraphics[width=85mm, angle =-0]{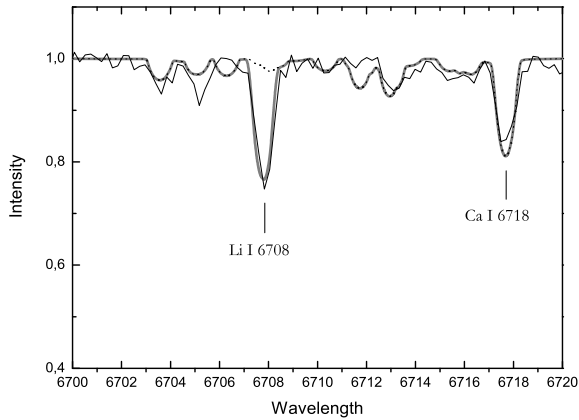}
%\vspace*{-3mm}
\caption{\label{Lit} The spectrum of RZ Psc around the Li 6708
\AA\, line (thin line). The synthetic spectra  with the Li excess
(100 times larger than the solar abundance)  (thick gray line) and
without excess (dashed line) are shown (see text for details).}
%\end{centering}
\end{figure}

Thus, the spectral analysis shows that the atmosphere of RZ Psc
has a huge excess of lithium (about 100 times the solar abundance)
and, according to this criterion, RZ Psc is not yet a main
sequence star. The considerably high (for this spectral type)
value of $v\,\sin{i}$ agrees with this conclusion (see the
discussion by Bouvier \cite{Bo}). On the other hand, in the
atmospheres of young solar-type stars (T Tauri stars) the excess
of lithium is usually much greater than in our case (see
Fig.~\ref{Ple}a ). For example, V718 Per (about the same spectral
type as RZ Psc) has a lithium excess equals to about 3.2-3.5 dex
(Grinin et al. \cite{Grin08}), which is indicative of a primordial
Li abundance (Pavlenko \& Magazzu \cite{Pav}). From this point of
view, RZ Psc is not any more a young star. To estimate its age, we
used the rough calibration of Li excess as a function of age for
clusters of different ages (King 1993; Soderblom et al.
\cite{Sod90,Sod93}; Sestito \& Randich 2005). Our analysis of
these data shows (see Fig.~\ref{Ple}) that the excess of Li in the
atmosphere of RZ Psc corresponds approximately to a stellar age
between the age of the Pleiades (about 70 Myr) and the Orion
(about 10 Myr) clusters. It is admittedly a rough estimate, taking
into account the high dispersion of Li abundance versus age in
both clusters. Nevertheless, this estimate qualitatively agrees
with the absence of IR excess in RZ Psc and its isolated position
far from the star-forming regions.

\begin{figure}
%\begin{centering}
%\label{2}
\includegraphics[width=85mm, angle =-0]{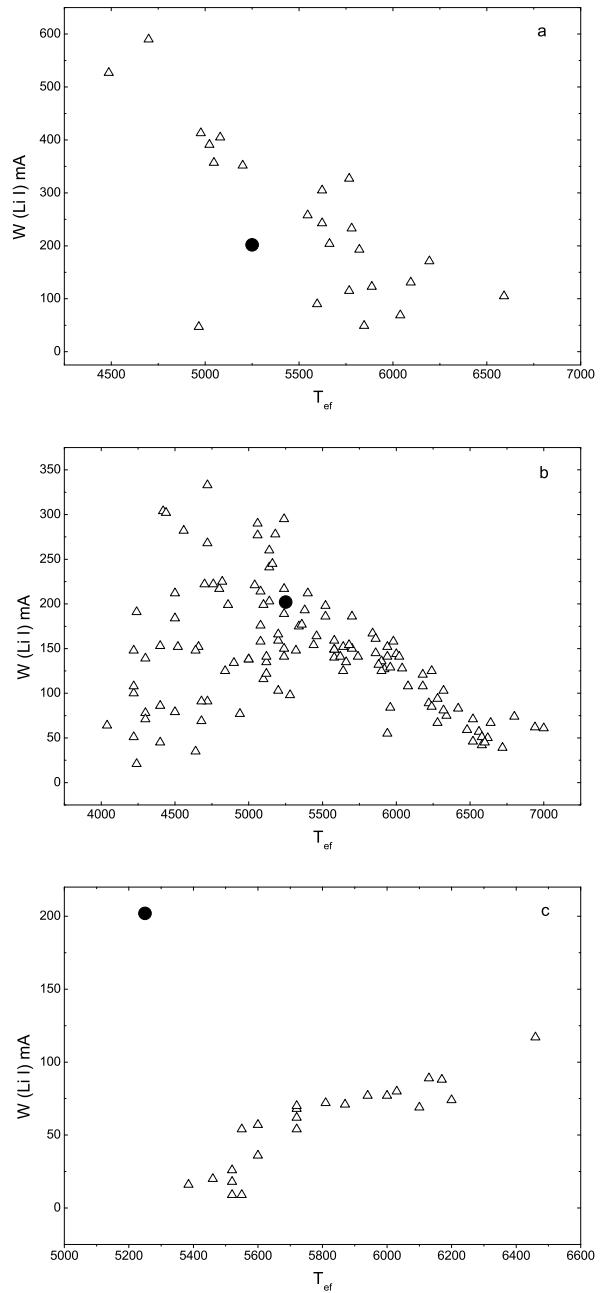}
%\vspace*{-3mm}
\caption{\label{Ple} Comparison of $EW$ (Li, 6708 \AA)  in the
spectrum of RZ Psc (circle) with the data for the stars in the
clusters of different age (triangles): Orion (10 Myr) from King
\cite{Kin}; Pleiades (70 Myr) and Hyades (600 Myr) from Soderblom
et al. \cite{Sod90,Sod93}.}
%\end{centering}
\end{figure}

\section{The kinematic age of RZ Psc}
Another possibility to estimate the age of RZ Psc is the proper
motion of the star. According to the Tycho-2 catalog by Hog et al.
\cite{Hog}, the proper motion of RZ Psc is quite large: pmRA =
25.4 $\pm$ 2 mas yr$^{-1}$, pmDE = -11.9 $\pm$ 2.1 mas yr$^{-1}$.
The corresponding values of the proper motion in the galactic
coordinate system are: pm$l = 7.89\pm 0.60$ deg. per Myr, pm$b$ =
-2.76$\pm 0.49$ deg. per Myr.

Assuming that RZ Psc has not moved away very far from its
birthplace near to the Galaxy plane (GP), one can calculate the
vertical component of its motion $W$ using the approximation by
Perrot \& Grenier \cite{Per} for the gravitation potential of the
Galactic disk. In this approximation the Galactic disk is locally
described by an infinite plane. The corresponding vertical
acceleration is given by
\begin{equation}\label{1}
    g(z) = -4\pi\textmd{G}\rho_d z_d (1 - e^{-z/z_d})\,,
\end{equation}
where $\textmd{G}$ is the gravitational constant, $\rho_d$ = 7.6 x
10$^{-2} M_\odot\, pc^{-3}$ is a volume density at $z$ = 0
(Cr\'ez\'e et al. \cite{Cre}), $z_d$ = 260$\pm$60 pc is an
exponential scale height (Ojha et al. \cite{Oj}).

Substituting $g(z)$ in the right part of equation of motion in z-
direction and replacing $dW/dt$ by $WdW/dz$, after integration on
$z$ we obtain the vertical component of the velocity

\begin{equation}\label{2}
    W(x) = \pm \sqrt{W_0^2 - v_d^2\,f(x)}\,,
\end{equation}
where $x = z/z_d$,
\begin{equation}\label{3}
     f(x) = x - 1 + e^{-x}\,,
\end{equation}
\begin{equation}\label{4}
    v_d = 2z_d(2\pi G\rho_d)^{1/2}\,,
\end{equation}
$W_0$ is the initial velocity at $x$ = 0. Signum "+" in Eq.
(\ref{2}) corresponds to the motion toward the southern Galactic
pole.

Equation~(\ref{2}) describes the oscillations of a star in the
gravitation field of the disk after ejection from the disk plane
with the initial velocity $W_0$: at some point $x_{m}$ the
velocity $W$ drops up to zero, whereupon the star goes back,
intersects the disk plane, and continues the motion at the
opposite side of the disk plane. With Eq. (\ref{2}) we can
calculate the time of the star motion from $x_1$ to $x_2$ ($x_2
\le$ $x_{m})$:

\begin{equation}\label{3}
    t(x_1,x_2) = t_d \int_{x_1}^{x_2}\frac{dx}{\sqrt{a^2 -
    f(x)}}\,,
\end{equation}
Here $a$ = $W_0/v_d$, $t_d$ = $z_d/v_d$ $\approx$ 12 Myr. The time
$t(0,x_m)$ = $p$/4, where $p$ is the period of oscillations of a
star in the gravity field of the disk. If the current value $W(x)
> 0$ (the star moves away the Galactic plane), then the
kinematic age $t_{k} = t(0,x)$. In the opposite case $t_{k} = p/4
+ t(x,x_m)$.

Accordingly, we need two parameters to calculate $t_k$: a) the
current distance $z$ of the star from the disk plane, and b) the
corresponding velocity $W(z)$. Both parameters depend on the
distance to the star $D$: i) $z = D\sin{b} - z_\odot$, where
$z_\odot$ is a distance of the Sun from the Galactic plane (in
pursuance Joshi \cite{Joshi} we adopt $z_\odot$ = 25 pc); ii)
$W(z) = V_r \sin{b} + D$ pm$b \cos{b} + W_\odot$, where pm$b$ is
the proper motion of RZ Psc in ster/sec, $W_\odot$ = -7 km/sec is
the velocity of the Sun toward the nordic Galactic pole (Perrot \&
Grenier 2003), $V_r = - 11.5\pm1.5$ km/sec - the radial velocity
of the star (Kaminski et al. 2000). For the estimate of $D$, we
used the photometric data for the out-of-eclipse state of RZ Psc
($V \approx 11.5^m $, Zaitseva \cite{Za85}) and adopted $T_{ef}$ =
5250 K, $\log{g}$ = 4.0 (see above), and the stellar radius $R_* =
1.5 R_\odot$ (which is compatible with $\log{g}$ = 4.0 at $M_*
\approx 1\,M_\odot$). The interstellar extinction toward RZ Psc
$A_V \approx$ 0 (Kaminski et al. 2000). Taking this into account
we obtain $D \approx$ 240 pc\footnote{At this distance, the proper
motion of RZ Psc would correspond to a tangential velocity
$\simeq$ 32 km\,s$^{-1}$}. Using this value at the calculations of
$z$, $z_m$ and $W(z)$ we obtain $t_k$ = 37.6$\pm 5.4$ Myr. Here,
the uncertainty is due to the uncertainty of pm$b$ (-2.76$\pm
0.49$ deg/Myr). About the same uncertainty appears at the
variations of $D$ within 240$\pm$50 pc: $t_k = 37.6^{-7.2}_{+6.9}$
Myr. Taking this into account we can give an approximate estimate
of the kinematic age of the star: $t_k \simeq$ 30-40
Myr\footnote{Note that the kinematic age of RZ Psc is comparable
with the age of the Gould Belt, a nearby starburst region where
many stars have formed over 30 to 40 Myr ago (see Perrot \&
Grenier \cite{Per}, and references there). This allows us to
assume that RZ Psc is a runaway member of this cluster.}.

Thus, both estimates of the age of RZ Psc, based on the lithium
excess and the proper motion of the star give values that
essentially exceed the characteristic time of dissipation of
protoplanetary disks ($\sim $ 10 Myr; Strom, Edwards \& Skrutskie
\cite{Strom}). This allows us to classify RZ Psc as an
intermediate object between MS stars with debris disks and UXOrs.

\section{Discussion and conclusion}
The question arises: if RZ Psc is a post UXOr, what is, in this
case, the reason for the quick and sporadic eclipses of the star?
The apparent answer would be a screening of the star from time to
time by material remnant from the CS disk: planetesimals, comets,
and rocks orbiting the star and dissipating in its nearest
vicinity.

The idea of comet-like activity in the neighborhoods of young
stars as a possible source of the variable CS extinction is not
new. It was claimed many years ago by Gahm \& Greenberg
\cite{Gahm}. However, in the case of young stars, there is a power
alternative source of the variable extinction: the inhomogeneous
matter of CS disks. This source dominates in UX Ori stars.
Therefore, the case of RZ Psc is interesting and important because
there are no alternative explanations for the rapid and sporadic
eclipses of the star.

Using the flux variation rate (about 0.1 mag. per hour) estimated
by Zaitseva \cite{Za85} from intense photometric monitoring of two
deep minima of RZ Psc, one can estimate the tangential velocity
$v_t$ of the opaque screen (the dust cloud) intersecting the
line-of-sight and its approximate distance to the star. Assuming
an stellar radius of $R_*$ = 1.5 $R_\odot$ and stellar mass of
$M_*$ = 1 $M_\odot $, we obtained $v_t \approx$ 40 km\,s$^{-1}$
and a distance $\approx$ 0.6 AU. In order to screen the star from
an observer during two days (the duration of eclipses), the clouds
have to be quite compact (about 0.05 AU).

The further study of RZ Psc and the search for stars with similar
properties can give valuable information about the disk properties
at the transitional phase of their evolution. It is important,
first of all, to observe the mid-IR excess of this star at
$\lambda \ge$ 5 $\mu$m. There are indications that this excess
exists, including the observations of the highly polarized
radiation in the deep minima (Kiselev, Minikulov \& Chernova
\cite{Kis}, Shakhovskoi, Grinin \& Rostopchina \cite{Shah}) and
the so-called blueing effect (Zaitseva 1985, Kardopolov, et al.
1980, Pugach 1981). Both these effects are caused in UX Ori stars
by the scattered radiation of CS dust (Grinin \cite{Grin88}). This
radiation is the relatively stabilized agent (different in
different stars), which provides the limitation of the amplitudes
of the algol-type minima. For RZ Psc ($\Delta V)_{max} \approx$
2.5, which corresponds to the intensity of scattered radiation
$I_{sc} \approx 0.1 I_*$,  (which is comparable with the level of
$I_{sc}$ in the other UXORs (Grinin et al. 1991). The absence of
near IR excess in RZ Psc means that in this case the main part of
CS dust lies far from the star, in the circumbinary disk. This
dust is probably formed (as in MS stars with debris disks) by
collisions of large particles, rocks, and planetesimals (see. e.g.
Wyatt \cite{Wyatt}) and is heated by stellar radiation. The energy
distribution of RZ Psc at mid-IR wavelengths will enable us to
estimate the characteristic size of the region occupied by CS
dust. The other important problem is to find the distance from the
star. It allows us to estimate more precisely the time when RZ Psc
has left its birthplace.

\begin{acknowledgements} We thank the staff of the Terskol Observatory
for the observations and the referee for the useful comments that
improved the paper. This research has been supported in part by
the program of the Presidium of RAS "Formation and Evolution of
Stars and Galaxies" and grant N.Sh.- 3645.2010.2.
\end{acknowledgements}


\begin{thebibliography}{}

%\bibliographystyle{aa}
%\bibliography{reference.bib}

\bibitem[2008]{Bo} Bouvier, J. 2008, A\&A, \textbf{489}, 53
\bibitem[1998]{Cre} Cr\'ez\'e, M., Chereul, E., Bienym\'e, O., Pichon, C. 1998, A\&A,
\textbf{329}, 920
\bibitem[2009]{daS} da Silva L., Torres, C.A.O, de la Reza R. et
al. 2009, A\&A, \textbf{508}, 833
\bibitem[2003]{Dull} Dullemond, C. P., van den Ancker, M. E., Acke, B. \& van Boekel,
R. 2003, ApJ, \textbf{594}, L47
\bibitem[1983]{Gahm} Gahm, G.H., \& Greenberg, J.M. 1983, in \emph{Asteroids, Comets, Meteors},
eds. by C.-I. Lagerkvist \& H. Rickman, Uppsala Univ. 375
\bibitem[1992]{Gal} Galazutdinov, G.A. 1992, SAO Preprint, N 92
\bibitem[1974]{GP74} Glass, I.S. \& Penston, M.V. 1974, MNRAS, \textbf{167}, 237
\bibitem[1988]{Grin88} Grinin, V.P. 1988, Soviet. Astron. Lett. \textbf{14}, 27
\bibitem[1991]{Grin91} Grinin, V.P., Kiselev, N.V., Minikulov, N.H, Chernova, G.P. \&
Voshchinnikov N.V. 1991, ApSS, \textbf{186}, 283
\bibitem[2008]{Grin08} Grinin, V., Stempels, H.C., Gahm, G.F., et al. 2008, A\&A, \textbf{489},
1233
\bibitem[1960]{Herbig} Herbig, G.H. 1960, ApJ, \textbf{131}, 632
\bibitem[2000]{Hog} Hog, E., Fabricius, C., Makarov, V. V. et al. 2000, A\&A, \textbf{355}, L27
\bibitem[2007]{Joshi} Joshi, Y.C. 2007, MNRAS, \textbf{378}, 768
\bibitem[2000]{Kam} Kaminskii, B.M., Kovalchuk, G.U. \& Pugach, A.F. 2000, Astronomy Reports,
\textbf{44}, 611
\bibitem[1980]{KSSh80} Kardopolov, V.I., Sahanionok, V.V. \& Shutemova, N.A. 1980, Perem.
Zvezdy, \textbf{21}, 310
\bibitem[1993]{Kin} King, J., 1993, AJ, \textbf{105}, 1087,
\bibitem[1991]{Kis} Kiselev, N.N., Minikulov, N.K. \& Chernova, G.P. 1991, Astrophysics,
\textbf{34}, 175
\bibitem[1999]{Kup} Kupka, F., Piskunov, N.E., Ryabchikova, T.A., Stempels, H.C.,
\& Weiss, W.W. 1999, A\&A, \textbf{138}, 119
\bibitem[1979]{Kur} Kurucz, R. 1979, ApJS, \textbf{40}, 1
\bibitem[2001]{Natta} Natta, A., Prusti, T., Neri, R., et al. 2001, A\&A, \textbf{371}, 186
\bibitem[1996]{Oj} Ojha, D.K., Bienym\'e, O., Robin, A.C. et al. 1996, A\&A, \textbf{311}, 456
\bibitem[1996]{Pav} Pavlenko, Ya.V. \& Magazzu, A. 1996, A\&A, \textbf{311}, 961
\bibitem[2003]{Per} Perrot, C.A. \& Grenier, I.A. 2003, A\&A, \textbf{404}, 519
\bibitem[1992]{Pis92} Piskunov, N.N. 1992, in \emph{Stellar Magnetizm}, eds. Yu.V. Glagolevsky
\& I.I.Romanjuk, (St. Petersburg, Nauka), 92
\bibitem[1995]{Pis95} Piskunov, N.E., Kupka, F., Ryabchikova, T.A., Weiss, W.W. \&
Jeffery, C.S. 1995, A\&A, \textbf{112}, 525
\bibitem[1981]{Pu81} Pugach, A.F. 1981, Astrophysics, \textbf{17}, 47
\bibitem[2005]{Ses} Sestito, P. \& Randich, S. 2005, A\&A, \textbf{442}, 615
\bibitem[2003]{Shah} Shakhovskoi, D.N., Grinin, V.P. \& Rostopchina, A.N. 2003, Astronomy Reports,
\textbf{47}, 580
\bibitem[1993]{Sod93} Soderblom, D.R., Jones, B.F., Balachandran, S., et al. 1993, AJ,
\textbf{106}, 1059
\bibitem[1990]{Sod90} Soderblom, D.R., Oey, M.S., Johnson, D.R.H., Stone, R.P.S.
1990, Astron. J., \textbf{99}, 595
\bibitem[1993]{Strom} Strom, S.E., Edwards, W. \& Skrutskie, M.F. 1993, in \emph{Protostars
and Planets} III, (ed. E.H. Levy \& J.I. Lunine), 837, (University
of Arizona Press, Tucson)
\bibitem[1996]{Wyatt} Wyatt, M. 2008, Ann. Rev. Astron. Astrophys. \textbf{46}, 339
\bibitem[1989]{Wen89} Wenzel, W. 1989, Inform. Bull. Var. Stars. \textbf{3280}, 1
\bibitem[1985]{Za85} Zaitseva, G.V. 1985, Variable Stars, \textbf{22}, 181

\end{thebibliography}
\end{document}